\documentclass[aps,amsmath,showpacs,amsfonts,10pt,twocolumn]{revtex4}
\usepackage{epsfig,graphicx}
\usepackage[english]{babel}
\usepackage{amsfonts}
\usepackage{amsmath}
\usepackage{latexsym}
\usepackage{graphics,bm}
\usepackage{natbib}
\usepackage{dcolumn}
\usepackage{bm}
\usepackage{rotating}

\begin{document}

\title{Phonon Laser Effect and Dicke-Hepp-Lieb Superradiant Phase Transition in Magnetic Cantilever Coupled to a Bose Einstein Condensate }

\author{ Aranya B Bhattacherjee$^{1}$ and Tobias Brandes$^{2}$}

\address{$^{1}$Department of Physics, A.R.S.D College, University of Delhi (South Campus), New Delhi-110021, India} \address{$^{2}$Institut f\"ur Theoretische Physik, TU Berlin, Germany}

\begin{abstract}
We propose a possibility of a phonon laser by coupling a Bose-Einstein condensate to a nanomechanical cantilever with a magnetic tip. Due to the magnetic coupling, atomic spin flips induce cantilever motion which can be used to produce a phonon laser. The system is described by the equivalent of the Jaynes-Cummings Hamiltonian. By controlling the number of atoms and the population inversion, one can obtain either a continuous wave (cw) or transient lasing. The two-body atom-atom interaction is also shown to coherently manipulate the lasing process. We also show that in the strong coupling limit, the same system can
undergo a Dicke-Hepp-Lieb superradiant phase transition. Exotic phase diagrams can be obtained by tuning the two body atom-atom interaction.
\end{abstract}


\maketitle

\section{Introduction}

There have been strong activities recently in quantum `optics' with phonons instead of photons, such
as experiments for the study and control of single phonons \cite{cleland}, or theoretical concepts of a phonon laser.

The phonon analog of the optical laser has been proposed in numerous physical systems so far.
To name a few, particular proposals are based on  paramagnetic ions in a lattice \citep{kittel}, paraelectric crystals \citep{vredevoe}, isolated trapped ions \citep{wallentowitz}, quantum wells \citep{lozovik}, semiconductors \citep{camps,liu,kabuss,okuyama}, nanomechanic systems \citep{bargatin}, nanomagnets \citep{chudnovsky}, and ultra-cold matter\citep{mendonca}. From the experimental point of view, phonon laser action has been demonstrated in cryogenic $\text{Al}_{2}\text{O}_{3}\text{:Cr}^{3+}$ \citep{tucker,hu,fokker}, $\text{Al}_{2}\text{O}_{3}\text{: V}^{4+}$ \citep{bron}, semiconductor superlattices \citep{kent}, harmonically bound magnesium ions \citep{vahala} and very recently in a compound microcavity system \citep{grudinin}.

Recently the field of cavity optomechanics has become an attractive research topic with Bose-Einstein condensate \citep{brennecke, murch, bhattacherjee09, bhattacherjee10, camerer, szirmai,hunger2,chen, chiara, steinke, chen2, zhang} and atomic ensembles \citep{singh,singh1,singh2,singh3,singh4,meiser}. A cavity opto-mechanical system, generally consists of an optical cavity with one movable end mirror. Such a system is utilized to cool a micromechanical resonator to its ground state by the pressure exerted by the cavity light field on the movable mirror. The studies on cavity opto-mechanics of atoms show that sufficiently strong and coherent coupling would enable studies of atom-oscillator entanglement, quantum state transfer, and quantum control of mechanical force sensors. Recent experiments have shown an impressive level of coherent control over micro- and nanomechanical oscillators. Magnetically coupling ultracold atoms to mechanical oscillators creates a unique setting where coherent quantum control over all degrees of freedom can be achieved \citep{treulein, hunger3}.

In this paper, we propose a phonon laser that operates like a two-level optical laser by  using a Bose-Einstein condensate of $^{87} \text{Rb}$ atoms magnetically coupled to a magnetic cantilever \citep{treulein}.
We also show that for strong coupling, the same system can undergo a  Dicke-Hepp-Lieb  type phase transition \cite{Hepp,emary,bhaseen} into a phonon superradiance regime. The two-body atom-atom interaction thereby plays a crucial role in the dynamics of the system.

\section{Phonon Lasing}

\begin{figure}[h]
\hspace{-0.0cm}
\begin{tabular}{c}
\includegraphics [scale=0.60]{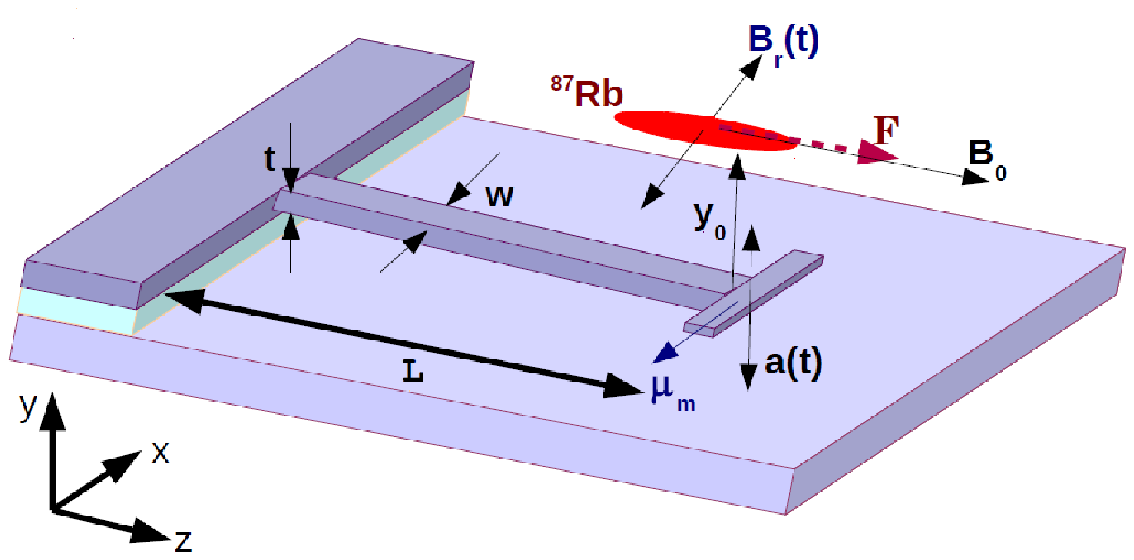}\\

\includegraphics [scale=0.55] {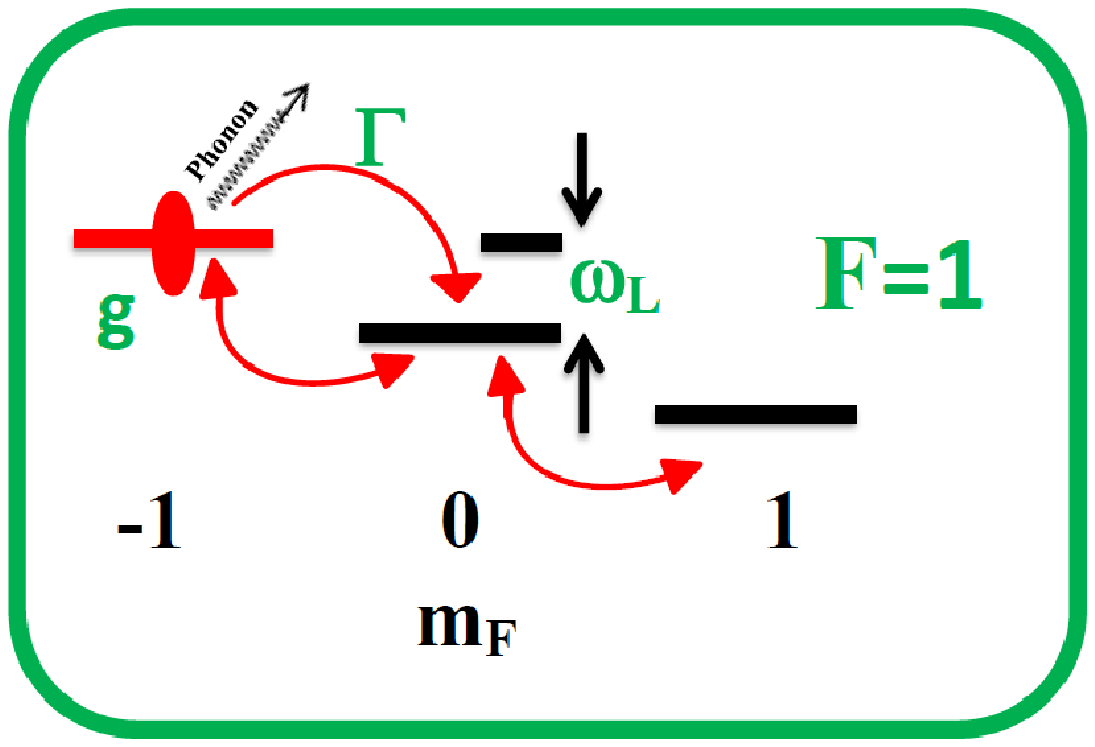}
 \end{tabular}
\caption{Top figure: Schematic view of the coupling of the BEC (in red) to the magnetic cantilever (blue). The BEC is at a distance $y_{0}$ from the cantilever. The cantilever performs out-of-plane mechanical oscillations denoted by $a(t)$. The oscillatory component of the magnetic field $B_{r}(t)$ couples the magnetic cantilever to the atomic spin $\textbf{F}$. Bottom figure: Hyperfine structure of $^{87}\text{Rb}$. Transition from the state $|F=1, m_{F}=-1\rangle$ to $|F=1, m_{F}=0\rangle$ leads to the emission of phonons.}
\label{f1}
\end{figure}

The phonon lasing device proposed here essentially consists of a gas of ultracold $^{87} \text{Rb}$ atoms at a distance $l_{0}$ above a cantilever resonator  \citep{treulein} with a ferromagnetic tip (fig.1), which creates a strong magnetic field $G_{m}=\frac{2 \mu_{0} |\vec{\mu}_{m}|}{4 \pi y_{0}^{4}}$ \citep{treulein}. Here, $\vec{\mu}_{m}$ is the magnetic moment of the ferromagnetic tip and $\mu_{0}$ is the permeability of free space.
The magnetic cantilever performs out-of-plane mechanical oscillations which transduces into an oscillatory magnetic field $\vec{B}_{r}(t)$. The atomic spin $\vec{F}$ interacts with $\vec{B}_{r}(t)$, where $\vec{\mu}=\mu_{B} g_{F} \vec{F}$ is the magnetic moment operator. The ground state hyperfine spin levels $|F=1,m_{F}\rangle$ of $^{87}\text{Rb}$ are also shown in Fig.1. The energy splitting between adjacent $m_{F}$ levels is given by the tunable Larmor frequency $\omega_{L}=\mu_{B} |g_{F}| B_{0}/\hbar$. The tunability of $\omega_{L}$ allows one to coherently control the detuning $\delta=\omega_{r}-\omega_{L}$, where $\omega_{r}$ is the mechanical frequency of the cantilever mode. Near resonance ($\delta \approx 0$), atom-phonon coupling leads to spin flips, between the ground level $|g\rangle=|F=1, m_{F}=0\rangle$ and the excited level $|e\rangle=|F=1, m_{F}=-1\rangle$. The transition $|g\rangle \Leftrightarrow |e\rangle$ can be decoupled from other $m_{F}$ levels \citep{treulein}.
This coupled system of BEC and magnetic cantilever (atoms driving the cantilever and the cantilever driving the atoms) leads to a kind of positive feedback that arises in all laser like system.
In order to observe phonon lasing, we must also introduce a pumping mechanism to compensate for the dissipation in the phonon number and the loss of condensate atoms. This can be achieved by pumping atoms in the excited state $|e\rangle$, so as to maintain a steady population inversion.

\subsection{Hamiltonian}
The coupled dynamics of the magnetic cantilever and the BEC can be described by the Hamiltonian,

\begin{equation}
H = H_{atom}+H_{phonon}+H_{atom-phonon}+H_{atom-atom},
\end{equation}

where,

\begin{eqnarray}
&H_{atom}& = \int d^{3}r  \psi_{g}^{\dagger}(r)  [  \frac{-\hbar^{2} \nabla^{2} }{2 m}+\hbar \omega_{g}+V_{g}(r)  ] \psi_{g}(r) \nonumber \\
&+& \int d^{3}r \psi_{e}^{\dagger}(r) [  \frac{-\hbar^{2} \nabla^{2} }{2 m}+\hbar \omega_{e}+V_{e}(r)  ] \psi_{e}(r),
\end{eqnarray}

\begin{equation}
H_{phonon}=\hbar \omega_{r} a^{\dagger}a,
\end{equation}

\begin{eqnarray}
H_{atom-phonon}&=& \hbar g (a^{\dagger}+a) [ \int \psi_{g}^{\dagger} (r) \psi_{e}(r)  d^{3}r \nonumber \\
&+& \int \psi_{e}^{\dagger}(r) \psi_{g}(r)  d^{3}r ],
\end{eqnarray}

\begin{eqnarray}
&H_{atom-atom}& = \sum_{i=g,e} \frac{2 \pi \hbar^{2} a_{ii}}{m}  \int d^{3}r  \   \psi_{i}^{\dagger}(r) \psi_{i}^{\dagger}(r) \psi_{i}(r) \psi_{i}(r) \nonumber \\
 &+&\frac{4 \pi \hbar^{2} a_{ge}}{m}  \int d^{3}r   \  \psi_{g}^{\dagger}(r)  \psi_{g}(r) \psi_{e}^{\dagger}(r) \psi_{e}(r).
\end{eqnarray}

Here, $\psi_{g}$ and $\psi_{e}$ are the ground and excited state wavefunction of the condensate. Also $\omega_{i}$ and $V_{i}(r)$ ( $i=g,e$ ) are the energies and the trapping potentials respectively for the ground and excited states of the condensate. $g$ is the atom-phonon coupling constant taken to be real. $m$ is the mass of single atom of the condensate and $a_{gg}$, $a_{ee}$ and $a_{ge}$ are the $s$-wave scattering lengths corresponding to  $ground-ground$, $excited-excited$ and $ground-excited$ atomic states respectively. Here we have taken $a_{eg}=a_{ge}$.

We now write \citep{steel}

\begin{equation}
\psi_{g}(r,t)= \sqrt{N} b_{0}(t)  \xi_{g}(r),
\end{equation}

\begin{equation}
\psi_{e}(r,t)= \sqrt{N} c_{0}(t) \xi_{e}(r),
\end{equation}
where $b_{0}(t)$ and $c_{0}(t)$ are the annihilation operators for the ground and excited state atoms, respectively.  Here, $\xi_{g}(r)$ and $\xi_{e}(r)$ are the single particle wave functions for the ground and excited state respectively satisfying the normalization $\sum_{i=g,e}\int d^{3}r \ |\xi_{i}(r)|^{2}=1$.
Ignoring counter rotating terms, we get the following second-quantized Hamiltonian in terms of the normalized operators, $b_{0}\rightarrow \sqrt{N} b_{0}$ and $c_{0}\rightarrow \sqrt{N} c_{0}$,

\begin{eqnarray}
H &=&\hbar \omega_{r} a^{\dagger} a + \hbar \omega_{0} b_{0}^{\dagger} b_{0}+ \hbar \omega_{1} c_{0}^{\dagger}c_{0} +\hbar [G a b_{0} c_{0}^{\dagger} + G^{*} a^{\dagger} c_{0} b_{0}^{\dagger}] \nonumber \\
&+& \frac{\hbar K_{gg}}{2} b_{0}^{\dagger}b_{0}^{\dagger}b_{0}b_{0}+\frac{\hbar K_{ee}}{2} c_{0}^{\dagger}c_{0}^{\dagger}c_{0}c_{0}+ \hbar K_{eg} b_{0}^{\dagger} b_{0} c_{0}^{\dagger} c_{0},
\end{eqnarray}
where

\begin{equation}
\hbar \omega_{0}= \int d^{3}r \ \xi^{*}_{g}(r) \left[ \frac{-\hbar^{2} \nabla^{2} }{2 m}+\hbar \omega_{g}+V_{g}(r) \right ] \xi_{g}(r),
\end{equation}

\begin{equation}
\hbar \omega_{1}= \int d^{3}r \ \xi^{*}_{e} (r) \left[ \frac{-\hbar^{2} \nabla^{2} }{2 m}+\hbar \omega_{e}+V_{ge}(r) \right ] \xi_{e}(r),
\end{equation}

\begin{equation}
\hbar K_{gg}= \frac{4 \pi \hbar^{2} a_{gg}}{m} \int d^{3}r \ |\xi_{g}(r)|^{4},
\end{equation}

\begin{equation}
\hbar K_{ee}= \frac{4 \pi \hbar^{2} a_{ee}}{m} \int d^{3}r \ |\xi_{e}(r)|^{4},
\end{equation}

\begin{equation}
\hbar K_{eg}= \frac{4 \pi \hbar^{2} a_{eg}}{m} \int d^{3}(r) \ |\xi_{g}(r)|^{2} |\xi_{e}|^{2},
\end{equation}

\begin{equation}
\hbar G= \hbar g \int d^{3}r \ \xi_{e}^{*}(r) \xi_{g}(r) .
\end{equation}

\subsection{Phonon laser mean field equations}
We now write down the Heisenberg equation of motion for the phonon operator $a$ and the atomic operators $b_{0}$ and $c_{0}$,

\begin{equation}
\dot{a}=-i \omega_{r} a -\frac{\Gamma}{2} -i G^{*}c_{0}b_{0}^{\dagger}
\end{equation}

\begin{eqnarray}
\dot{b_{0}}^{\dagger}&=& i [\omega_{0} b_{0}^{\dagger}+a G c_{0}^{\dagger}+ \frac{K_{gg}}{2}|b_{0}|^{2} b_{0}^{\dagger} \nonumber \\
&+&K_{eg} |c_{0}|^{2} b_{0}^{\dagger}] -\frac{\gamma}{2} b_{0}^{\dagger},
\end{eqnarray}

\begin{eqnarray}
\dot{c_{0}} &=& -i [\omega_{1} c_{0} + a G c_{0} + \frac{K_{ee}}{2}|c_{0}|^{2} c_{0}^{\dagger} \nonumber \\
&+& K_{eg} |b_{0}|^{2} c_{0}^{\dagger}] -\frac{\gamma}{2} c_{0}.
\end{eqnarray}

Here, $\Gamma$ and $\gamma$ are the damping rates for the phonons and the atoms, respectively. Now in terms of the polarization $p=b_{0}^{\dagger} c_{0}/N$ and population inversion $\Delta n= (|c_{0}|^{2}-|b_{0}|^{2})/N$, the Heisenberg equations of motion can be rewritten in the rotating frame of the phonon frequency $\omega_{r}$ as,

\begin{equation}
\dot{a}= -\frac{\Gamma}{2} a + G^{*} N p,
\end{equation}

\begin{equation}
\dot{p}= -i \Delta \omega p - \frac{\gamma}{2} p+ G a \Delta n,
\end{equation}

\begin{equation}
\dot{\Delta n}= \frac{\gamma}{2} (\Delta n_{eq}- \Delta n)- 2 [G^{*} a^{\dagger}p+ G a p^{*}].
\end{equation}
Here, we have taken $G\rightarrow i G$ and $\Delta \omega=\omega_{L}- \omega_{r}+ \frac{K_{-}}{4}-(\frac{K_{+}}{4}-K_{eg})N \Delta n$, $K_{-}=K_{gg}-K_{ee}$, $K_{+}=K_{gg}+K_{ee}$. Also $\Delta n_{eq}$ is the equilibrium value of $\Delta n$.

After factorization, the mean-field
steady state solutions of Eqns. (18)-(20) lead to a critical number of atoms $N_{cr}$ required to support a continuous wave (cw) laser, $N>N_{cr}=\Gamma (\Delta \omega^{2}+\gamma^{2}/4)/(|G|^{2}\Delta n_{eq} \gamma)$. For possible experimental values mentioned \citep{treulein}, $N_{cr}=300/\Delta n_{eq}$. For $\Delta n_{eq}=0.2$, $N_{cr}=1.5 \times 10^{3}$ atoms which is a reasonable number.

\subsection{Transient solutions}
One of the important predictions of the Jaynes-Cummings model \citep{jaynes} 
are coherent population oscillations between an oscillator and a (pseudo) spin, i.e. a two-level system.  Such energy oscillations can also be observed in our current system if the effective frequency of oscillations, $\sqrt{|\Delta n_{eq}|N}G$, is larger than the fastest relaxation of the system $\gamma$.
From this condition, we can estimate the minimum number of atoms $N_{t}$ required to observe this transient phenomena as $N_{t}=\gamma^{2}/(|\Delta n_{eq}|G^{2})$ $\approx$ $1.25 \times 10^{4}$. On the other hand if $N_{t}>N>N_{cr}$, a single pulse can be produced instead of energy oscillations between the cantilever and the atoms.

The two transient outputs mentioned above are shown in Fig. 2 by numerical integration of Eqns.(18)-(20). In Fig. 2(a) for $N=1 \times 10^{5}$ (thin line), coherent energy oscillations are observed while for $N=4 \times 10^{3}$ (thick line), one single pulse is obtained which takes away a large part of the energy at one go. If the number of atoms $N$ becomes less than $N_{cr}$, the single pulse disappears completely. The tails of the output pulses decay as $\Gamma^{-1}$. Fig. 2(b) shows the influence of the two-body interaction on the transients. Keeping the number of atoms $N=1 \times 10^{5}$ fixed, increasing the value of $(K_{+}/4-K_{eg})$ decreases the amplitude of the coherent oscillations but increases the frequency of oscillations. Atomic two-body interactions can be manipulated by Feshbach resonances \citep{pethick}.

\begin{figure}[h]
\hspace{-0.0cm}
\begin{tabular}{cc}
\includegraphics [scale=0.60]{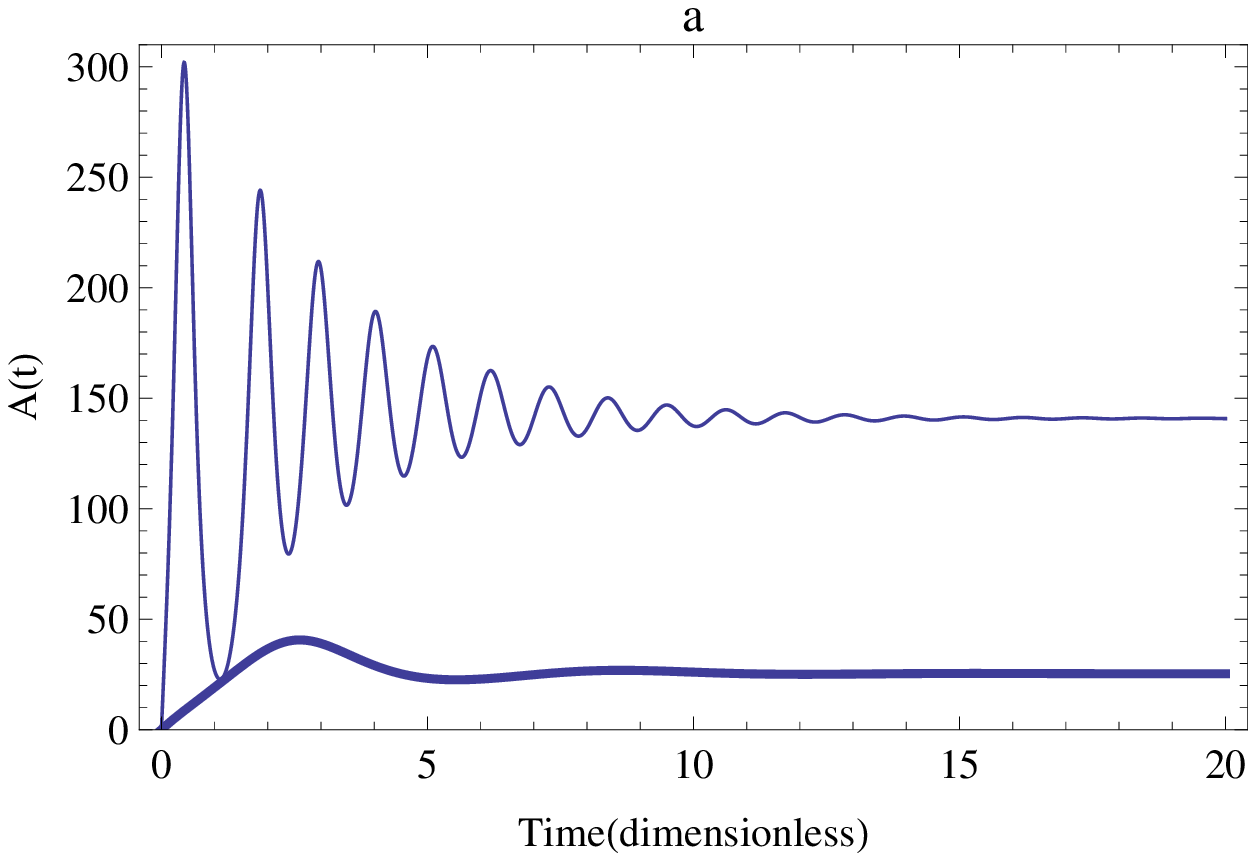}\\

\includegraphics [scale=0.60] {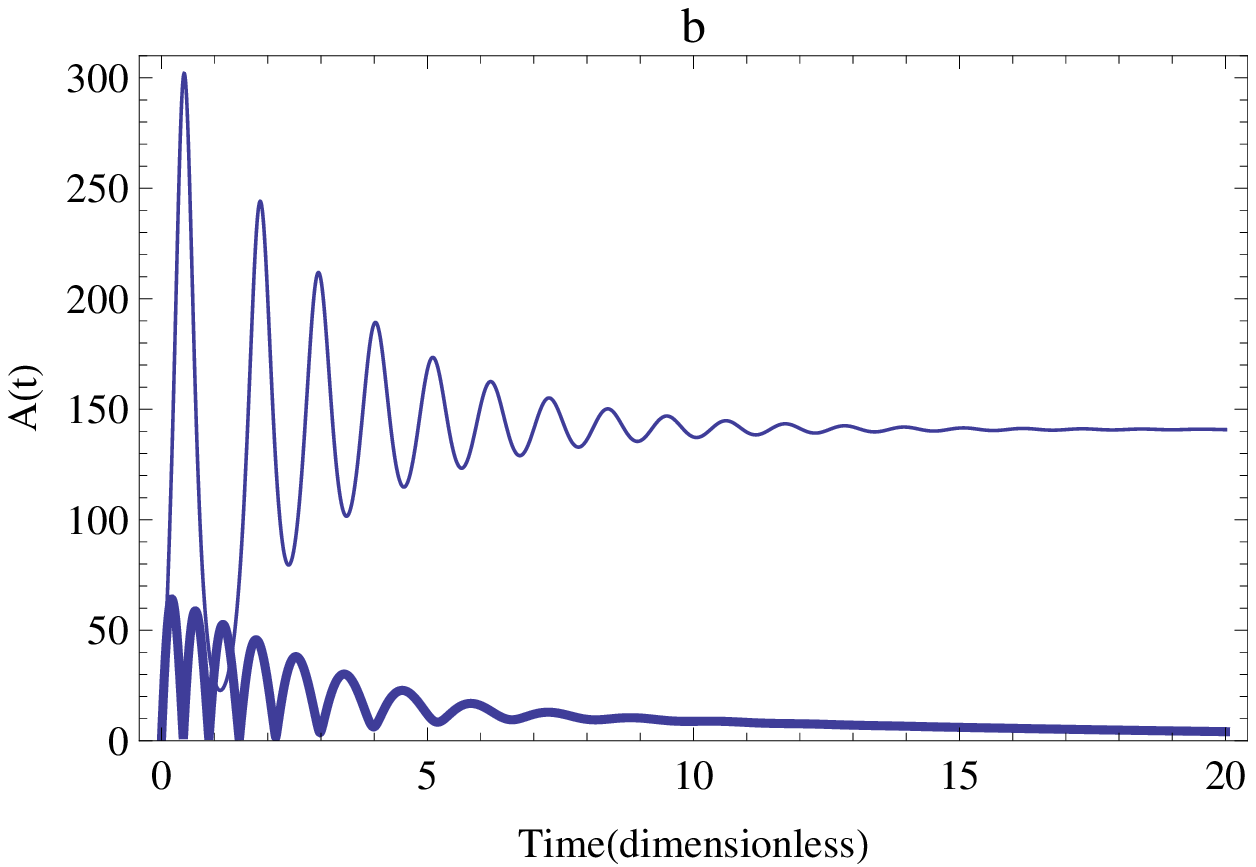}\\
 \end{tabular}
\caption{Plot (a): Plot of $A(t)=a^{\dagger}(t) a(t)$ as a function of time for $N=1 \times 10^{5}$ (thin line), $N=4 \times 10^{3}$ (thick line), Plot (b): Plot of $A(t)=a^{\dagger}(t) a(t)$ as a function of time for $(\frac{K_{+}}{4}-K_{eg})=0$ (thin line), $(\frac{K_{+}}{4}-K_{eg})=2.2 \times 10^{-3}$ (thick line). All parameters are dimensionless with respect to the atomic damping rate $\gamma$. Other parameters used are: $\Gamma/\gamma=0.5$, $G/\gamma=0.02$, $\omega_{L}-\omega_{r}=0$, $\Delta n_{eq}=0.2$.}
\label{f2}
\end{figure}

\section{Phonon Superradiance Phase Transition}

We now demonstrate that the system considered here is capable of producing phonon superradiance. The BEC of $N$ two-level atoms with level spacing $\hbar \omega_{L}$ can be described by a collective spin $S=N/2$. In the weak atom-phonon coupling limit, the counter-rotating terms $a^{\dagger} S_{+}$ and $a S_{-}$ in the usual Dicke Hamiltonian are usually neglected (rotating wave approximation ) \citep{emary}. In the strong atom-phonon coupling regime and including atom-atom interactions, one can show that the Hamiltonian of Eqn.(8) can be rewritten as,

\begin{eqnarray}\label{dicke1}
H &=& \hbar \omega_{r} a^{\dagger}a + \hbar (\omega_{L}+\gamma_{0}) S_{z}+\hbar \gamma_{1}S^{2}_{z} \nonumber \\
&+& \hbar \lambda (a^{\dagger}+a)(S_{+}+S_{-}),
\end{eqnarray}

where $S_{z}$, $S_{-}$ and $S_{+}$ are the collective spin operators of the BEC \citep{treulein}. $S_{\pm}=S_{x} \pm i S_{y}$ and $S_{x}=\sqrt{2} g_{F} F_{x}$ \citep{treulein} ($F_{x}$ is the $x$ component of the atomic spin). Also, $\lambda=G_{N}/\sqrt{2S}$, $G_{N}=G \sqrt{N}$, $\gamma_{0}=(K_{ee}-K_{gg})/4$ and $\gamma_{1}=(K_{gg}+K_{ee}-2 K_{eg})/2$. Note the nonlinear term $\hbar \gamma_{1} S_{z}^{2}$ ($S_{z}^{2}\neq 1$) is expected to show rich collective dynamics and phase diagrams due to the fact that $\gamma_{1}$ can be both positive and negative. We now study the dynamics arising out of the Hamiltonian of Eqn.(\ref{dicke1}). The equations of motion for $a$, $S_{z}$ and $S_{-}$ are given by

\begin{equation}
\dot{a}=-i \omega_{r} a-\kappa a-i \lambda (S_{+}+S_{-}),
\end{equation}

\begin{equation}
\dot{S}_{z}=\lambda (a^{\dagger}+a)(S_{-}-S_{+}),
\end{equation}

\begin{eqnarray}
\dot{S}_{-}&=& -i (\omega_{L}+\gamma_{0}+\gamma_{1}) S_{-}-2 i \gamma_{1} S_{-}S_{z} \nonumber \\
&+& i 2 \lambda (a^{\dagger}+a) S_{z}.
\end{eqnarray}

Here we have taken the phonon damping rate $\Gamma/2=\kappa$.

\subsection{Mean field solutions}
We first study the long time,
mean-field steady state solutions with $\dot{\vec{S}}=0$ and $\dot{a}=0$. Following \citep{bhaseen}, we identify two stable states; the normal state ($\Downarrow$, denoted by $N$ in the phase diagrams), with all spins pointing down, $S_{z}=-N/2$ and no phonons, $a=0$,  and the inverted state ($\Uparrow$, denoted by $I$ in the phase diagrams) with all spins pointing up, $S_{z}=N/2$ and no phonons.

We now look for other interesting configurations by analyzing the steady states. Writing $a=a_{1}+i a_{2}$ ,$S_{\pm}=S_{x}\pm i S_{y}$, $\dot{a}=0$, $\dot{S}_{z}=0$ and $\dot{S}_{-}=0$ leads to
\begin{equation} \label{c1}
[\kappa+i \omega_{r}]a=-i \lambda S_{x},
\end{equation}

\begin{equation}\label{c2}
\omega_{0} S_{x}=-2 \gamma_{1} S_{z} S_{x}+4 \lambda a_{1} S_{z},
\end{equation}

\begin{equation}\label{c3}
(2 \gamma_{1} S_{z}+\omega_{0})S_{y}=0,
\end{equation}
where $\omega_{0}=\omega_{L}+\gamma_{0}+\gamma_{1}$. From Eqn.(\ref{c3}), we observe that there are two classes of solutions depending on whether $S_{y}=0$ or $S_{z}=-\omega_{0}/2 \gamma_{1}$. $S_{y}=0$ is the usual superradiant phase in the Dicke model. For $S_{y}=0$, we obtain the steady state population difference,

\begin{equation}
S_{z}=\frac{\omega_{0}(\kappa^{2}+\omega_{r}^{2})}{8 \lambda^{2} \omega_{r}-2 \gamma_{1}(\kappa^{2}+\omega_{r}^{2})}.
\end{equation}
The critical coupling strength for the onset of superradiance is obtained by setting $\vec{S}=(0,0,\pm N/2)$. One obtains,

\begin{equation}\label{crit1}
\lambda^{2} N_{\pm}=\pm \frac{(\omega_{0}\pm \gamma_{1} N)(\kappa^{2}+\omega_{r}^{2})}{4 \omega_{r}}.
\end{equation}

For $\gamma_{1}=0$, $\gamma_{0}=0$,
we have
$\lambda^{2} N_{\pm}=\pm \frac{\omega_{L}(\omega_{r}^{2}+\kappa^{2})}{4 \omega_{r}}$, which is the usual Dicke model critical point \citep{emary}. For $\gamma_{1}\neq 0$, we find that the two body atom-atom interactions turn out to be a convenient and new handle to tune the critical point. The second solution $S_{z}=-\omega_{0}/2\gamma_{1}$ leads to $a=0$, which turns out to be the normal phase.

\subsection{Fluctuations}
We now discuss the instability of the normal phase and the inverted phase by considering fluctuations of phonon number and the spin $S_{-}$ around the steady state. To this end, we write $a\rightarrow a+\delta a$ and $S_{-}=S_{-}+\delta S_{-}$, where $a=0$, $S_{-}=0$ and $S_{z}=\mp N/2$. We consequently obtain the linearized equations

\begin{equation}
\dot{\delta a}=-(\kappa+i \omega_{r}) \\\ \delta a-i \lambda (\delta S_{-}+\delta S_{+}),
\end{equation}

\begin{equation}
\dot{\delta S_{-}}=-i \tilde{\omega_{0}}_{\mp} \\\ \delta S_{-} \mp i \lambda N (\delta a+\delta a^{\dagger}),
\end{equation}
where $\tilde{\omega_{0}}_{\mp}=\omega_{0}\mp \gamma_{1} N$. We look for solutions of the form $\delta a=A e^{-i \eta t}+B^{*} e^{i \eta^{*}t}$ and $\delta S_{-}= C e^{-i \eta t}+D^{*} e^{i \eta^{*}t}$ and equating coefficients with same time dependence, one obtains equations for $A$, $B$, $C$ and $D$. Following \citep{bhaseen}, $\eta$ satisfies,

\begin{eqnarray}\label{d1}
(\eta^{2}-\omega_{r}^{2}-\kappa^{2})(\eta^{2} &-& \tilde{\omega_{0}}_{\mp}^{2}) \mp  4 \lambda^{2}N \omega_{r} \tilde{\omega_{0}}_{\mp} \nonumber \\
&-& i2 \kappa \eta (\eta^{2}- \tilde{\omega_{0}}_{\mp})=0.
\end{eqnarray}

The boundary between unstable (exponentially growing ) and stable (exponentially decaying) solutions corresponding to Eqn. (\ref{d1}) having real solutions for $\eta$. The imaginary part of Eqn.(\ref{d1}) vanishes when $\eta=0$ or $\eta=\tilde{\omega_{0}}_{\mp}$. $\eta=\tilde{\omega_{0}}_{\mp}$ implies that the real part of Eqn.(\ref{d1}) vanishes when $\tilde{\omega_{0}}_{\mp}=0$. This implies both the normal and inverted states become unstable when $\omega_{0}=\pm \gamma_{1} N$. For $\eta=0$, the real part of Eqn. (\ref{d1}) becomes zero when $(\omega_{r}^{2}+\kappa^{2})\tilde{\omega_{0}}_{\mp}^{2} \mp 4 \lambda^{2} N \omega_{r} \tilde{\omega_{0}}_{\mp}=0$. This gives the same condition as Eqn.(\ref{crit1}). Consequently this implies that the onset of the superradiant phase is accompanied by the instability of the normal($N$) and inverted phase ($I$).

The dynamical phase diagram corresponding to the dynamics of Eqns. (30,31), can be calculated by the corresponding eigenvalues. In this non-equilibrium setting, the eigenvalues are given by,
\begin{equation}
\omega_{\mp}= \frac{\pm 2 \lambda^{2} N}{\omega_{0} \mp \gamma_{1} N} \pm \sqrt{\frac{4 \lambda^{4}N^{2}}{(\omega_{0}\mp \gamma_{1}N)^{2}}-\kappa^{2}}.
\end{equation}

\subsection{Phase diagrams}

The dynamical phase diagrams emerging from the eigenvalues of Eqn.(33) are shown in Fig.3. For $\gamma_{1}N=0$ and for the positive eigenvalue $\omega_{-}$, the phase diagram of Fig. 3(a) reflects the equilibrium phase diagram of the Dicke model, having a transition from the normal phase (N) at low $\lambda \sqrt{N}$ to the superradiant normal phase (SRN) at higher value of $\lambda \sqrt{N}$. This dynamical phase transition occurs at $\lambda \sqrt{N}=\dfrac{\omega_{L}(\omega_{r}^{2}+\kappa^{2})}{4 \omega_{r}}$. As $\omega_{-} \rightarrow 0$, the critical value of $\lambda \sqrt{N}$ required for superradiance tends to infinity. Also we find as in \citep{bhaseen}, for negative eigenvalues $\omega_{+}$, this open dynamical system shows signature of non-equilibrium dynamics; the normal state ($\Downarrow$) becomes unstable and an inverted state ($S_{z}=N/2$, $a=0$, denoted by $I$) and superradiant inverted phase (denoted by $SRI$) emerges which is a stable state. The inverted state is the mirror image of the normal state which is also reflected in the equation of motion Eqns.(22)-(24), which have an inversion symmetry for $\omega_{r}\rightarrow -\omega_{r}$, $\vec{S} \rightarrow - \vec{S}$, $a \rightarrow a^{*}$ and $\gamma_{1}=0$.

In the presence of a finite $\gamma_{1}$, this inversion symmetry is immediately broken which is evident from the Eqns.(22)-(24). Fig.3(b) and Fig.3(c) illustrates this broken inversion symmetry. For $\gamma_{1}N=-0.04$ (Fig.3(b)), the phase boundary between the $N$ phase and $SRN$ phase recedes to higher $\lambda \sqrt{N}$ values while the phase boundary connecting the $I$ phase and the $SRI$ phase shifts towards lower $\lambda \sqrt{N}$ values. Exactly the opposite is true for $\gamma_{1} N=0.04$ (Fig.3(c)) A further increase in the values of $\gamma_{1}N= \pm 0.06$ gives rise to regions where both the $SRN$ and $SRI$ phases coexist (Fig.3(d) and Fig.3(e)). The influence of increasing the phonon damping rate ($\kappa$) is shown in Fig.3(f). An increase in the phonon damping rate separates the $SRN$ and $SRI$ phases further i.e the regions of $N$ phase and $I$ phases increases.

\begin{figure}[h]
\hspace{-0.0cm}
\begin{tabular}{cc}
\includegraphics [scale=0.35]{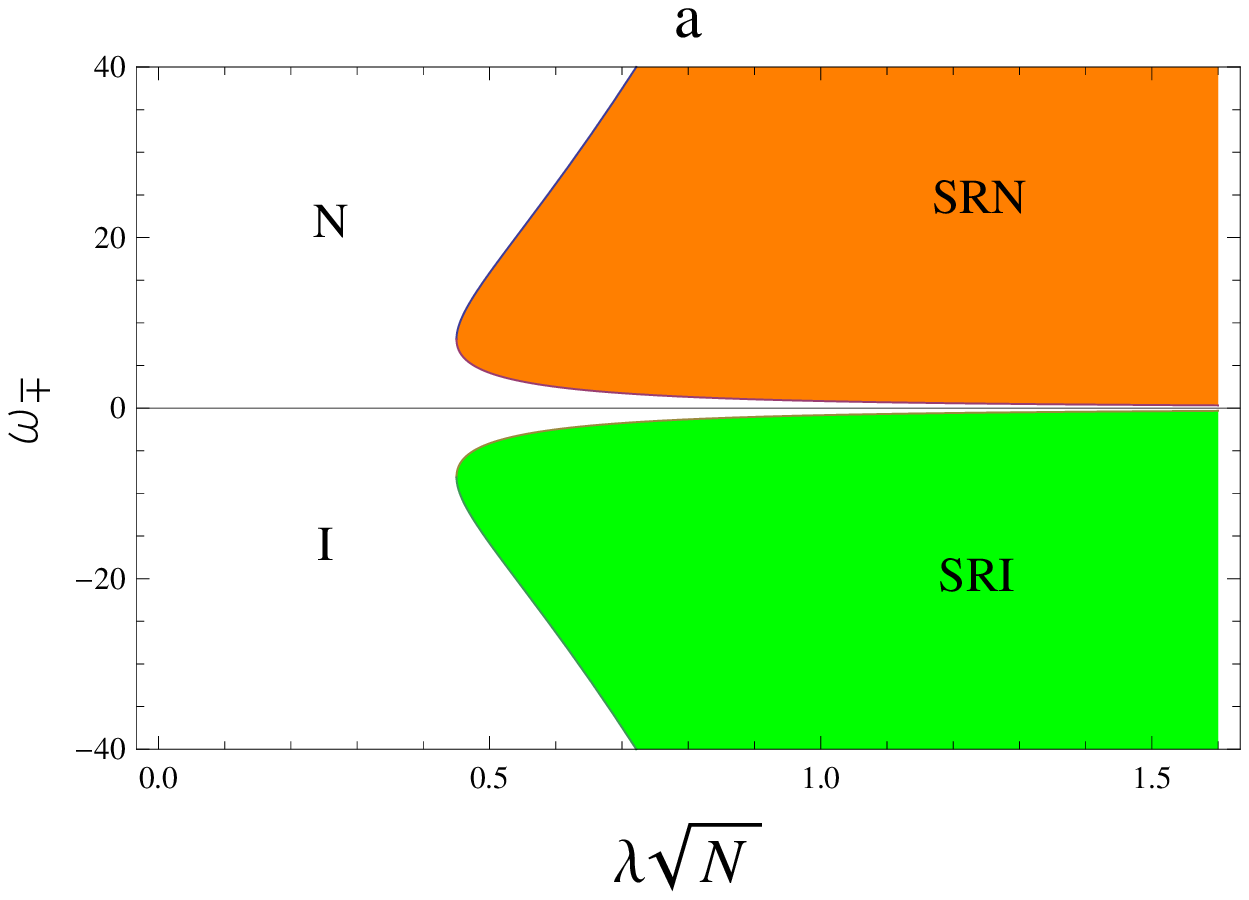}&\includegraphics [scale=0.35] {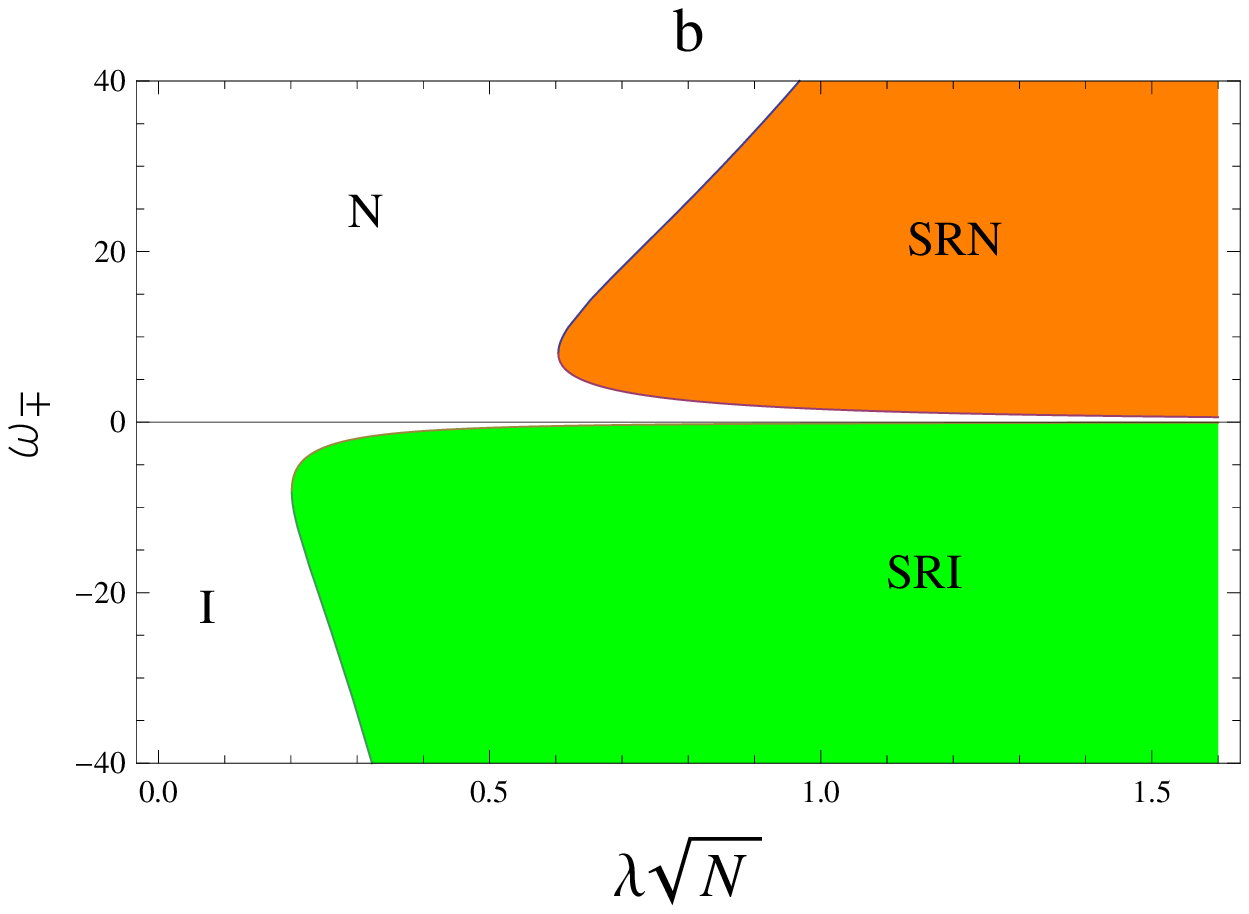}\\
\includegraphics [scale=0.35]{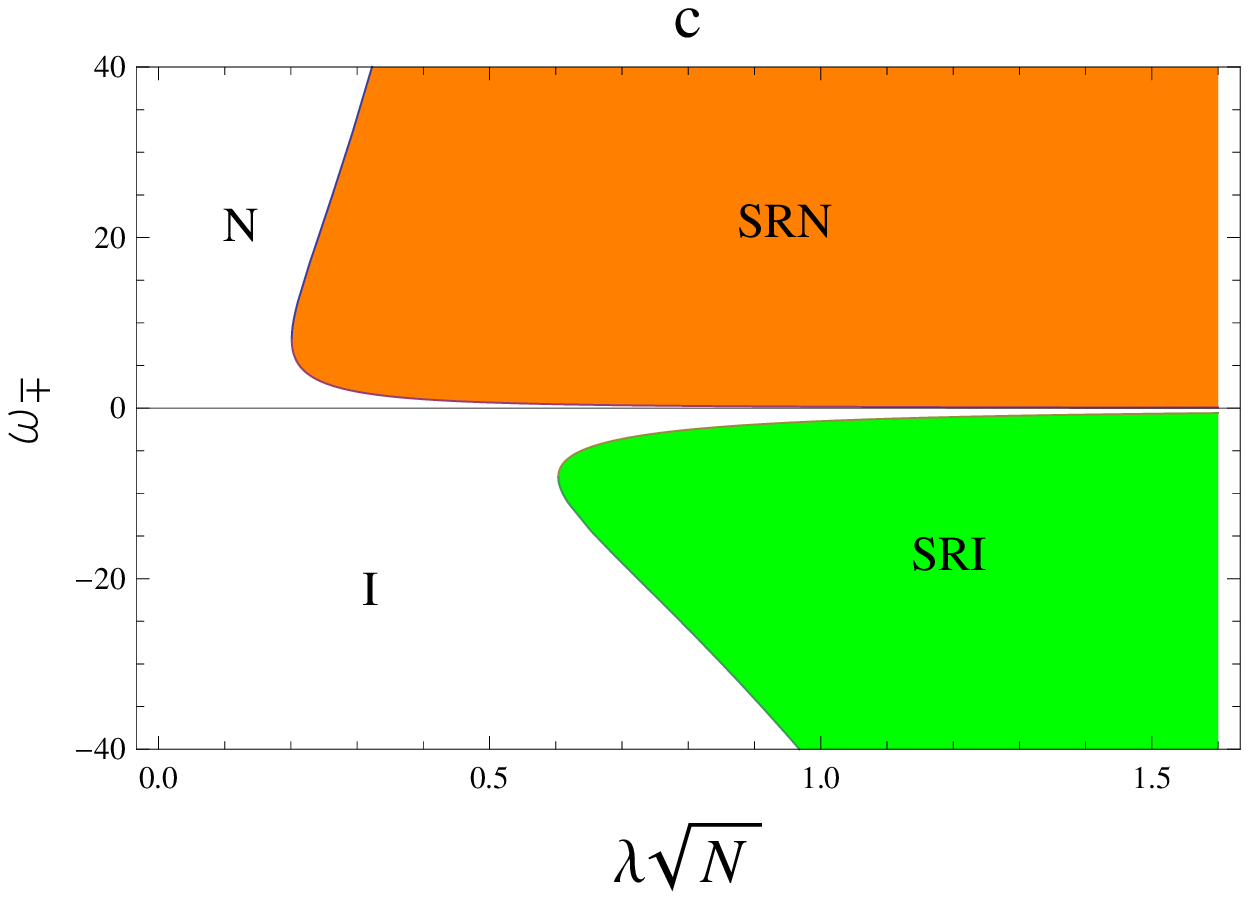}& \includegraphics [scale=0.35] {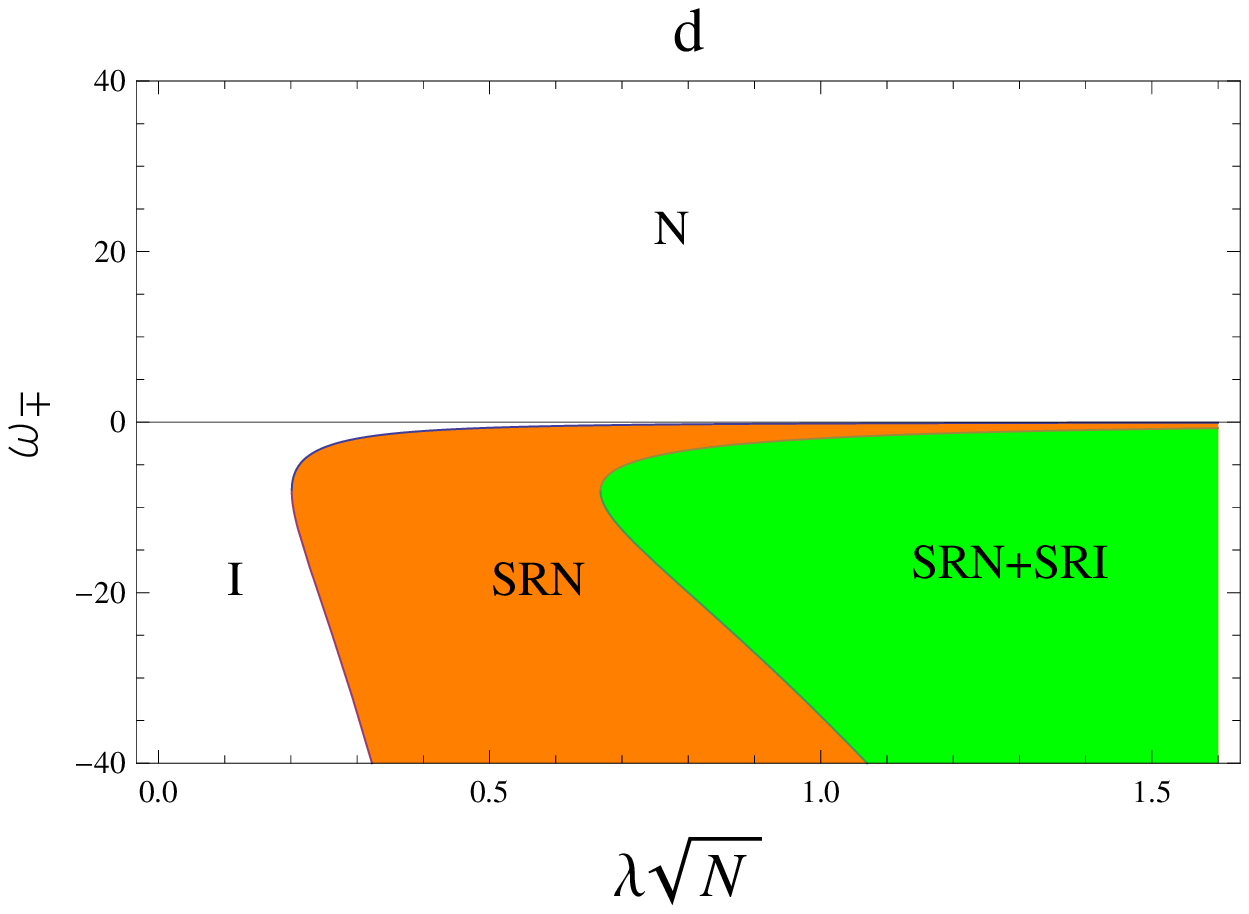}\\
\includegraphics [scale=0.35] {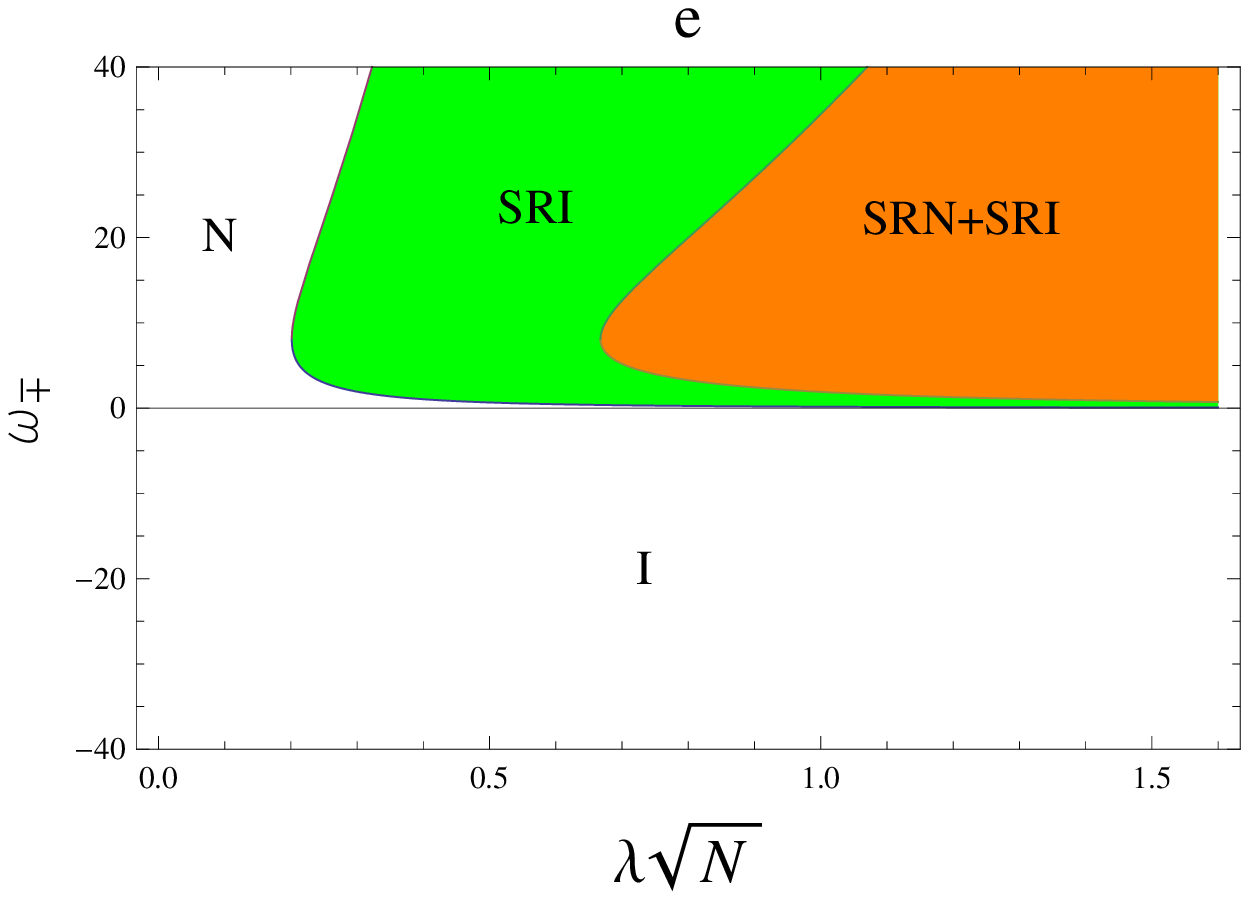}& \includegraphics [scale=0.35] {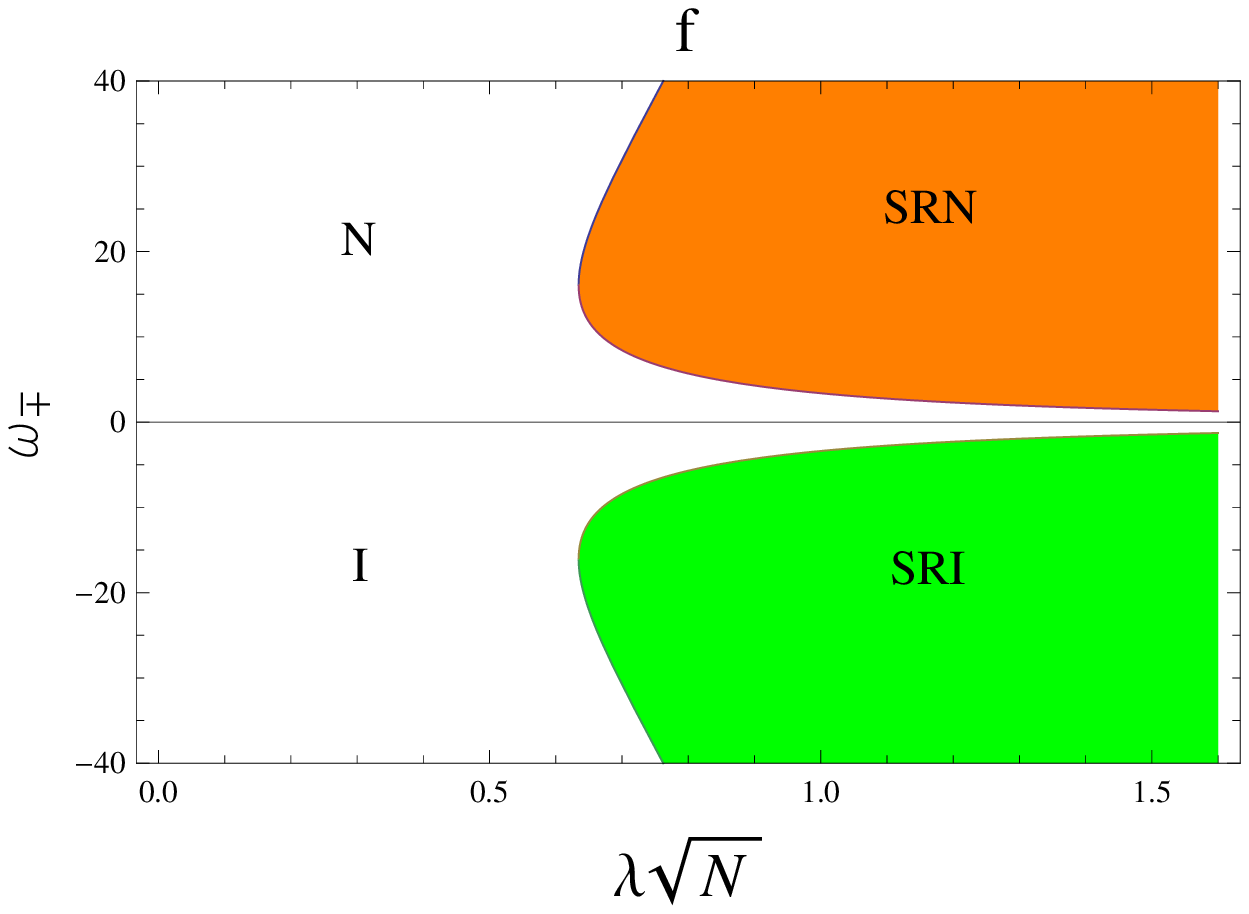}
 \end{tabular}
\caption{Dynamical phase diagrams depicting the various phases (normal phase $N$, inverted phase $I$, superradiant normal phase $SRN$ and superradiant inverted phase $SRI$). Plot (a): $\gamma_{1}N=0$,$\kappa=8.1$ , Plot (b): $\gamma_{1}N=-0.04$,  Plot (c): $\gamma_{1}N=0.04$,  Plot (d): $\gamma_{1}N=0.06$, Plot (e): $\gamma_{1}N=-0.06$ , Plot (f): $\gamma_{1} N=0$, $\kappa=16.1$}
\label{f3}
\end{figure}

\section{Conclusions}
In conclusion, we have shown that a phonon laser can be fabricated based on magnetically coupling a ultra-cold atomic cloud to the mechanical oscillations of a nanoscale magnetic cantilever in close analogy to a two-level optical laser system. By controlling the number of atoms, one can switch between solitary pulses and transient pulses. The transients can also be controlled by the atomic two body interaction. We have also demonstrated that the system considered here is capable of producing phonon superradiance. For large atom-mechanical mode coupling, the system can be described by the Dicke type Hamiltonian. The two-body interaction gives rise to a nonlinear term proportional to $S_{z}^{2}$, which gives rise to some interesting phase diagrams. By appropriately tuning the two-body interaction, we get regions in the phase diagram where the superradiant normal and superradiant inverted phase coexist.

\section{Acknowledgements}
 A. Bhattacherjee acknowledges financial support from the Department of Science and Technology, New Delhi for financial assistance vide grant SR/S2/LOP-0034/2010. T Brandes acknowledges  support via DFG Grants No. BR1528/7-1, No. 1528/8-1, No. SFB 910, and No. GRK 1558.

\end{document}